\def\preflag{y}
\nopagenumbers
%
%
%
\def\unredoffs{} \def\redoffs{\voffset=-.31truein\hoffset=-.59truein}
\def\speclscape{}
%
%
%
%
\magnification=1200
\hsize=15.1truecm
\vsize=23.5truecm 
\def\prepr{y}
\def\header{\global\voffset=1truecm\advance\vsize by-\voffset
            \headline=
            {\ifnum\pageno>1 
               \ifx\prepr\preflag
                 \vbox{\hsize=15truecm\advance\hsize by -2pt\hfill}
               \else
                 \ifodd\pageno
                   \vbox{\hsize=15truecm\advance\hsize by -2pt
                    \centerline{\runtitle}}
                 \else
                   \vbox{\hsize=15truecm\advance\hsize by -2pt
                    \centerline{\runauth}}
                 \fi
               \fi
             \else 
               \ifx\preflag\prepr
                 \vbox{\hsize=15.0truecm
\centerline{\hbox{\vbox{\baselineskip 0.20truein\noindent\anbf To 
Appear in the {\anbfsl Proceedings of the
16$^{\hbox{\anbfsls th}}$ International Liquid Crystal Conference}, 
24-28 June 1996,
Kent, OH, USA ({\anbfsl Mol. Cryst. Liq. Cryst.})}}}}
               \fi
             \fi}}
\newbox\leftpage \newdimen\fullhsize \newdimen\hstitle \newdimen\hsbody
\tolerance=1000\hfuzz=2pt
\catcode`\@=11 
\def\bigans{b }
\def\answ{b }
%

\ifx\answ\bigans\message{(This will come out unreduced.}
\unredoffs\baselineskip=.24truein plus 2pt minus 1pt
\hsbody=\hsize \hstitle=\hsize 
\else\message{(This will be reduced.} \let\l@r=L
\magnification=1000\baselineskip=16pt plus 2pt minus 1pt \vsize=7truein
\redoffs \hstitle=8truein\hsbody=4.75truein\fullhsize=10truein\hsize=\hsbody
\output={\ifnum\pageno=0 
  \shipout\vbox{\speclscape{\hsize\fullhsize\makeheadline}
   \hbox to \fullhsize{\hfill\pagebody\hfill}}\advancepageno
  \else
 \almostshipout{\leftline{\vbox{\pagebody\makefootline}}}\advancepageno 
  \fi}
\def\almostshipout#1{\if L\l@r \count1=1 \message{[\the\count0.\the\count1]}
      \global\setbox\leftpage=#1 \global\let\l@r=R
 \else \count1=2
  \shipout\vbox{\speclscape{\hsize\fullhsize\makeheadline}
      \hbox to\fullhsize{\box\leftpage\hfil#1}}  \global\let\l@r=L\fi}
\fi
%
\newcount\yearltd\yearltd=\year\advance\yearltd by -1900

\ifx\preflag\prepr\footline={\hss\tenrm\folio\hss}\else\footline={\hss}\fi
%

\def\draftmode{\message{ DRAFTMODE }\def\draftdate{{\rm preliminary draft:
\number\month/\number\day/\number\yearltd\ \ \hourmin}}%
\headline={\hfil\draftdate}\writelabels\baselineskip=20pt plus 2pt minus 2pt
 {\count255=\time\divide\count255 by 60 \xdef\hourmin{\number\count255}
  \multiply\count255 by-60\advance\count255 by\time
  \xdef\hourmin{\hourmin:\ifnum\count255<10 0\fi\the\count255}}}
\def\nolabels{\def\wrlabeL##1{}\def\eqlabeL##1{}\def\reflabeL##1{}}
\def\writelabels{\def\wrlabeL##1{\leavevmode\vadjust{\rlap{\smash%
{\line{{\escapechar=` \hfill\rlap{\sevenrm\hskip.03in\string##1}}}}}}}%
\def\eqlabeL##1{{\escapechar-1\rlap{\sevenrm\hskip.05in\string##1}}}%
\def\reflabeL##1{\noexpand\llap{\noexpand\sevenrm\string\string\string##1}}}
\nolabels
%
\global\newcount\secno \global\secno=0
\global\newcount\meqno \global\meqno=1
\def\newsec#1{\global\advance\secno by1\message{(\the\secno. #1)}

\global\subsecno=0\eqnres@t\noindent{\the\secno. #1}
\writetoca{{\secsym} {#1}}\par\nobreak\medskip\nobreak}
\def\eqnres@t{\xdef\secsym{\the\secno.}\global\meqno=1\bigbreak\bigskip}
\def\sequentialequations{\def\eqnres@t{\bigbreak}}\xdef\secsym{}
\global\newcount\subsecno \global\subsecno=0
\def\subsec#1{\global\advance\subsecno by1\message{(\secsym\the\subsecno. #1)}
\ifnum\lastpenalty>9000\else\bigbreak\fi
\noindent{\sl\secsym\the\subsecno. #1}\writetoca{\string\quad 
{\secsym\the\subsecno.} {#1}}\par\nobreak\medskip\nobreak}

\def\appendix#1#2{\global\meqno=1\global\subsecno=0\xdef\secsym{\hbox{#1.}}
\bigbreak\bigskip\noindent{\bf Appendix #1. #2}\message{(#1. #2)}
\writetoca{Appendix {#1.} {#2}}\par\nobreak\medskip\nobreak}
%
%
\def\eqnn#1{\xdef #1{(\secsym\the\meqno)}\writedef{#1\leftbracket#1}%
\global\advance\meqno by1\wrlabeL#1}
\def\eqna#1{\xdef #1##1{\hbox{$(\secsym\the\meqno##1)$}}
\writedef{#1\numbersign1\leftbracket#1{\numbersign1}}%
\global\advance\meqno by1\wrlabeL{#1$\{\}$}}
\def\eqn#1#2{\xdef #1{(\secsym\the\meqno)}\writedef{#1\leftbracket#1}%
\global\advance\meqno by1$$#2\eqno#1\eqlabeL#1$$}
%
\newskip\footskip\footskip14pt plus 1pt minus 1pt 
\def\footnotefont{\ninepoint}\def\f@t#1{\footnotefont #1\@foot}
\def\f@@t{\baselineskip\footskip\bgroup\footnotefont\aftergroup\@foot\let\next}
\setbox\strutbox=\hbox{\vrule height9.5pt depth4.5pt width0pt}
\global\newcount\ftno \global\ftno=0
\def\foot{\global\advance\ftno by1\footnote{$^{\the\ftno}$}}
%
\newwrite\ftfile   
\def\footend{\def\foot{\global\advance\ftno by1\chardef\wfile=\ftfile
$^{\the\ftno}$\ifnum\ftno=1\immediate\openout\ftfile=foots.tmp\fi%
\immediate\write\ftfile{\noexpand\smallskip%
\noexpand\item{f\the\ftno:\ }\pctsign}\findarg}%
\def\footatend{\vfill\eject\immediate\closeout\ftfile{\parindent=20pt
\centerline{\bf Footnotes}\nobreak\bigskip\input foots.tmp }}}
\def\footatend{}
%
%
\global\newcount\refno \global\refno=1
\newwrite\rfile
\def\ref{$^{\the\refno}$\nref}
\def\nref#1{\xdef#1{$^{\the\refno}$}\writedef{#1\leftbracket#1}%
\ifnum\refno=1\immediate\openout\rfile=refs.tmp\fi
\chardef\wfile=\rfile\immediate
\write\rfile{\noexpand\item{{\the\refno}.\ }\reflabeL{#1\hskip.31in}\pctsign}\global\advance\refno by1\findarg}
\def\findarg#1#{\begingroup\obeylines\newlinechar=`\^^M\pass@rg}
{\obeylines\gdef\pass@rg#1{\writ@line\relax #1^^M\hbox{}^^M}%
\gdef\writ@line#1^^M{\expandafter\toks0\expandafter{\striprel@x #1}%
\edef\next{\the\toks0}\ifx\next\em@rk\let\next=\endgroup\else\ifx\next\empty%
\else\immediate\write\wfile{\the\toks0}\fi\let\next=\writ@line\fi\next\relax}}
\def\striprel@x#1{} \def\em@rk{\hbox{}} 
\def\lref{\begingroup\obeylines\lr@f}
\def\lr@f#1#2{\gdef#1{\ref#1{#2}}\endgroup\unskip}

\def\addref#1{\immediate\write\rfile{\noexpand\item{}#1}} 
\def\startrefs#1{\immediate\openout\rfile=refs.tmp\refno=#1}
\def\xref{\expandafter\xr@f}\def\xr@f[#1]{#1}
\def\refs#1{\count255=1$^{\r@fs #1{\hbox{}}}$}
\def\r@fs#1{\ifx\und@fined#1\message{reflabel \string#1 is undefined.}%
\nref#1{need to supply reference \string#1.}\fi%
\vphantom{\hphantom{#1}}\edef\next{#1}\ifx\next\em@rk\def\next{}%
\else\ifx\next#1\ifodd\count255\relax\xref#1\count255=0\fi%
\else#1\count255=1\fi\let\next=\r@fs\fi\next}
%

%
\newwrite\ffile\global\newcount\figno \global\figno=1
\def\fig{Figure~\the\figno\nfig}
\def\nfig#1{\xdef#1{Figure~\the\figno}%
\writedef{#1\leftbracket fig.\noexpand~\the\figno}%
\ifnum\figno=1\immediate\openout\ffile=figs.tmp\fi\chardef\wfile=\ffile%
\immediate\write\ffile{\noexpand\medskip\noexpand\item{Fig.\ \the\figno. }
\reflabeL{#1\hskip.55in}\pctsign}\global\advance\figno by1\findarg}
\def\xfig{\expandafter\xf@g}\def\xf@g fig.\penalty\@M\ {}
\def\figs#1{figs.~\f@gs #1{\hbox{}}}
\def\f@gs#1{\edef\next{#1}\ifx\next\em@rk\def\next{}\else
\ifx\next#1\xfig #1\else#1\fi\let\next=\f@gs\fi\next}
\newwrite\lfile
{\escapechar-1\xdef\pctsign{\string\%}\xdef\leftbracket{\string\{}
\xdef\rightbracket{\string\}}\xdef\numbersign{\string\#}}

\def\writestop{\def\writestoppt{\immediate\write\lfile{\string\pageno%
\the\pageno\string\startrefs\leftbracket\the\refno\rightbracket%
\string\def\string\secsym\leftbracket\secsym\rightbracket%
\string\secno\the\secno\string\meqno\the\meqno}\immediate\closeout\lfile}}
\def\writestoppt{}\def\writedef#1{}
\def\seclab#1{\xdef #1{\the\secno}\writedef{#1\leftbracket#1}\wrlabeL{#1=#1}}
\def\subseclab#1{\xdef #1{\secsym\the\subsecno}%
\writedef{#1\leftbracket#1}\wrlabeL{#1=#1}}
\newwrite\tfile \def\writetoca#1{}
\def\leaderfill{\leaders\hbox to 1em{\hss.\hss}\hfill}
\def\writetoc{\immediate\openout\tfile=toc.tmp 
   \def\writetoca##1{{\edef\next{\write\tfile{\noindent ##1 
   \string\leaderfill {\noexpand\number\pageno} \par}}\next}}}

%
\catcode`\@=12 
%
\edef\tfontsize{\ifx\answ\bigans scaled\magstep2\else scaled\magstep4\fi}
 \tfontsize  \tfontsize
 \tfontsize \font\titlei=cmmi10 \tfontsize
\font\titleis=cmmi7 \tfontsize \font\titleiss=cmmi5 \tfontsize
\font\titlesy=cmsy10 \tfontsize \font\titlesys=cmsy7 \tfontsize
\font\titlesyss=cmsy5 \tfontsize  \tfontsize
 \tfontsize  \tfontsize
 \tfontsize 
\font\anbfsl=cmbxsl10 scaled\magstep1 \font\anbfsls=cmbxsl10
\font\anbf=cmbx10 scaled\magstep1
\skewchar\titlei='177 \skewchar\titleis='177 \skewchar\titleiss='177
\skewchar\titlesy='60 \skewchar\titlesys='60 \skewchar\titlesyss='60
 \ifx\answ\bigans\else scaled\magstep1\fi
\ifx\answ\bigans\else

 \font\absi=cmmi10 scaled\magstep1
\font\absis=cmmi7 scaled\magstep1 \font\absiss=cmmi5 scaled\magstep1
\font\abssy=cmsy10 scaled\magstep1 \font\abssys=cmsy7 scaled\magstep1
\font\abssyss=cmsy5 scaled\magstep1 
\skewchar\absi='177 \skewchar\absis='177 \skewchar\absiss='177
\skewchar\abssy='60 \skewchar\abssys='60 \skewchar\abssyss='60
\fi
\font\ninerm=cmr9 \font\sixrm=cmr6 \font\ninei=cmmi9 \font\sixi=cmmi6 
\font\ninesy=cmsy9 \font\sixsy=cmsy6 \font\ninebf=cmbx9 
\font\nineit=cmti9 \font\ninesl=cmsl9 \skewchar\ninei='177
\skewchar\sixi='177 \skewchar\ninesy='60 \skewchar\sixsy='60 
\def\ninepoint{\def\rm{\fam0\ninerm}
\textfont0=\ninerm \scriptfont0=\sixrm \scriptscriptfont0=\fiverm
\textfont1=\ninei \scriptfont1=\sixi \scriptscriptfont1=\fivei
\textfont2=\ninesy \scriptfont2=\sixsy \scriptscriptfont2=\fivesy
\textfont\itfam=\ninei \def\it{\fam\itfam\nineit}\def\sl{\fam\slfam\ninesl}%
\textfont\bffam=\ninebf \def\bf{\fam\bffam\ninebf}\rm} 
%
%

\hyphenation{anom-aly anom-alies coun-ter-term coun-ter-terms}
\def\inv{^{\raise.15ex\hbox{${\scriptscriptstyle -}$}\kern-.05em 1}}

\def\Dsl{\,\raise.15ex\hbox{/}\mkern-13.5mu D} 
\def\dsl{\raise.15ex\hbox{/}\kern-.57em\partial}

\def\lspace{\ifx\answ\bigans{}\else\qquad\fi}
\def\lbspace{\ifx\answ\bigans{}\else\hskip-.2in\fi} 
\def\boxeqn#1{\vcenter{\vbox{\hrule\hbox{\vrule\kern3pt\vbox{\kern3pt
	\hbox{${\displaystyle #1}$}\kern3pt}\kern3pt\vrule}\hrule}}}
\def\mbox#1#2{\vcenter{\hrule \hbox{\vrule height#2in
		\kern#1in \vrule} \hrule}}  
%

\def\darr#1{\raise1.5ex\hbox{$\leftrightarrow$}\mkern-16.5mu #1}

\def\roughly#1{\raise.3ex\hbox{$#1$\kern-.75em\lower1ex\hbox{$\sim$}}}

\input epsf
\header
\def\div{\nabla\!\cdot\!}
\def\dot{\!\cdot\!}
\def\nb{{\bf\hat n}}
\def\Rb{{\bf R}}
\def\Rbh{{\bf\hat R}}
\def\rb{{\bf r}}
\def\curl{\nabla\times}

\def\runauth{\hskip 0.17truecm RANDALL D. KAMIEN}
\def\runtitle{\hskip 0.17truecm WEAK CHIRALITY IN ORDERED DNA PHASES}

\def\nudge{$^{\!,\!}$}

\def\tiny{\scriptscriptstyle\rm}

\def\kbT{k_{\scriptscriptstyle\rm B}T}

\def\bo#1{{\cal O}(#1)}

\def\up#1{\leavevmode \raise.16ex\hbox{#1}}

\def\free{\hbox{$\cal F$}}
\def\bold#1{\setbox0=\hbox{$#1$}%
     \kern-.010em\copy0\kern-\wd0
     \kern.025em\copy0\kern-\wd0
     \kern-.020em\raise.0200em\box0 }
\def\angstrom{\hbox{\AA}}

\lref\DG{P.G.~de~Gennes, \underbar{Solid State Commun.}, \underbar{14}, 997
(1973).}

\lref\TER{E.M.~Terentjev, \underbar{Europhys. Lett.}, \underbar{23}, 27
(1993).}

\lref\TON{J.~Toner, \underbar{Phys. Rev. A}, \underbar{27}, 1157 (1983).}

\lref\GIA{C.~Gianessi, \underbar{Phys. Rev. A}, \underbar{28}, 350 (1983);
\underbar{Phys. Rev. A}, \underbar{34}, 705 (1986).}

\lref\LIVO{
F.~Livolant, \underbar{Physica A}, \underbar{176}, 117 (1981).}

\lref\HN{B.I.~Halperin and D.R.~Nelson, \underbar{Phys. Rev. Lett.},
\underbar{41}, 121 (1978);
D.R.~Nelson
and B.I.~Halperin, \underbar{Phys. Rev. B}, \underbar{19}, 2457 (1979).}

\lref\PN{P.~Nelson and T.~Powers, \underbar{Phys. Rev. Lett.},
\underbar{69}, 3409 (1992);
\underbar{J. Phys. II (Paris)}, \underbar{3}, 1535 (1993).}

\lref\TGB{S.R.~Renn and T.C.~Lubensky, \underbar{Phys. Rev. A},
\underbar{38}, 2132 (1988); T.C.~Lubensky
and S.R.~Renn, \underbar{Phys. Rev. A},
\underbar{41}, 4392 (1990).}

\lref\MEYER{R.B.~Meyer, \underbar{Appl. Phys. Lett.},
\underbar{12}, 281 (1968); \underbar{Appl. Phys. Lett.}, \underbar{14}, 208
(1969).}

\lref\KT{J.~Toner, \underbar{Phys. Rev. Lett.}, \underbar{68}, 1331 (1992);
R.D.~Kamien and J.~Toner, \underbar{Phys. Rev. Lett.}, \underbar{74}, 3181
(1995).}

\lref\FEL{L.G.~Fel, \underbar{Phys. Rev. E}, \underbar{52}, 702 (1995).}

\lref\HKL{A.B.~Harris, R.D.~Kamien and T.C.~Lubensky, {\sl in preparation}
(1996).}

\lref\LCBO{R.D.~Kamien, \underbar{J. Phys. II \up(Paris\up)}, \underbar{6}, 461
(1996).}

\lref\KN{R.D.~Kamien and D.R.~Nelson, \underbar{Phys. Rev. Lett.},
\underbar{74}, 2499 (1995);
\underbar{Phys. Rev.} \underbar{E}, \underbar{53}, 650 (1996).}

\def\oneskip{\vskip 0.166truein}
\settabs\+aaaaa&&\cr
\pageno=1
\voffset 0.8truecm
\hoffset 0.7truecm
{}~
\oneskip\oneskip\oneskip
{\+&WEAK CHIRALITY IN ORDERED DNA PHASES\cr}
\oneskip\oneskip\oneskip
{\+&RANDALL D. KAMIEN\cr}
\oneskip\oneskip
{
\baselineskip=0.16truein
{\+&Department of Physics and Astronomy, University of Pennsylvania,\cr}
{\+&Philadelphia, PA 19104, USA\cr}
\oneskip\oneskip\oneskip
}
\+&\hbox{\vbox{\baselineskip 0.16truein\hsize=13.1truecm
\noindent\underbar{Abstract}
Recent experiments\ref\NIH{R.~Podgornik, H.H.~Strey, K.~Gawrisch, D.C.~Rau,
A.~Rupprecht and V.A.~Parsegian, \underbar{Proc. Nat. Acad. Sci.} {\sl in
press} (1996).}
on aligned DNA show hexatic order with no sign of macroscopic chirality.
I make the analogy between smectic liquid crystals
and chiral hexatics and show how the absence
of chirality cannot occur
in a thermodynamic phase of chiral molecules.  In addition, I
discuss the microscopic origin of chiral mesophases in liquid crystals and
show that, within the context of central forces between ``atoms'' on
``molecules'',
chiral interactions can occur only if there are biaxial correlations
between the mesogens.  Weak biaxial correlations can therefore lead to small
cholesteric pitches. }}\cr
\oneskip
\ifx\preflag\prepr\centerline{(5 June 1996)}\else\oneskip\fi\oneskip
\noindent\underbar{INTRODUCTION}\oneskip

Experimental realizations of the liquid crystalline hexatic, a
phase with broken orientational but not translational order\HN, are
difficult to find\ref\rdata{For a review, see
K.J.~Standburg, \underbar{Rev. Mod. Phys.}, \underbar{60}, 161 (1988).}.
More exotic hexatic structures have been proposed\TON\ in three dimensions
in which there is nematic order and, simultaneously, hexatic order in the
plane perpendicular to the director $\nb$.  Until recently, only the
liquid state (the regular nematic) and the hexagonal state
(the hexagonal columnar phase) had been observed, typically in long, chiral
molecules such as DNA\LIVO.  In fact, the liquid-like phase of
chiral polymers is cholesteric -- the nematic phase will always start
to twist in the presence of chirality.  The hexagonal columnar phase, however,
need not twist: there is a thermodynamic region of elastic constants
in which twist is expelled.  The twist can come in either via a
tilt-grain-boundary
state in which the nematic order twists or via a moir\'e state in which the
polymers
braid about each other and the local crystalline axes rotate along
the polymer axis\KN.

In a new set of experiments\NIH\ a phase with both nematic and hexatic
order was found in DNA, in addition to a cholesteric phase at lower
concentration.
In the first section of this talk I will argue that this is rather surprising:
in an ``N+6'' phase Landau theory predicts that, unless a microscopic
parameter is precisely tuned, either the nematic order or the
bond-orientational
order must twist.  If that were the case, the X-ray scattering in the plane
perpendicular to the nematic director would be a powder average over
many different, rotated hexatic regions.  Thus one might expect that there
should
be a ring in the $q_{\tiny\perp}$ plane, rather than the observed $\cos
6\theta$ modulation
as shown in Figure 1.
\bigskip
\vskip 0.2truein
\centerline{\vbox{\epsfxsize=4truein\hsize=4truein\epsfbox{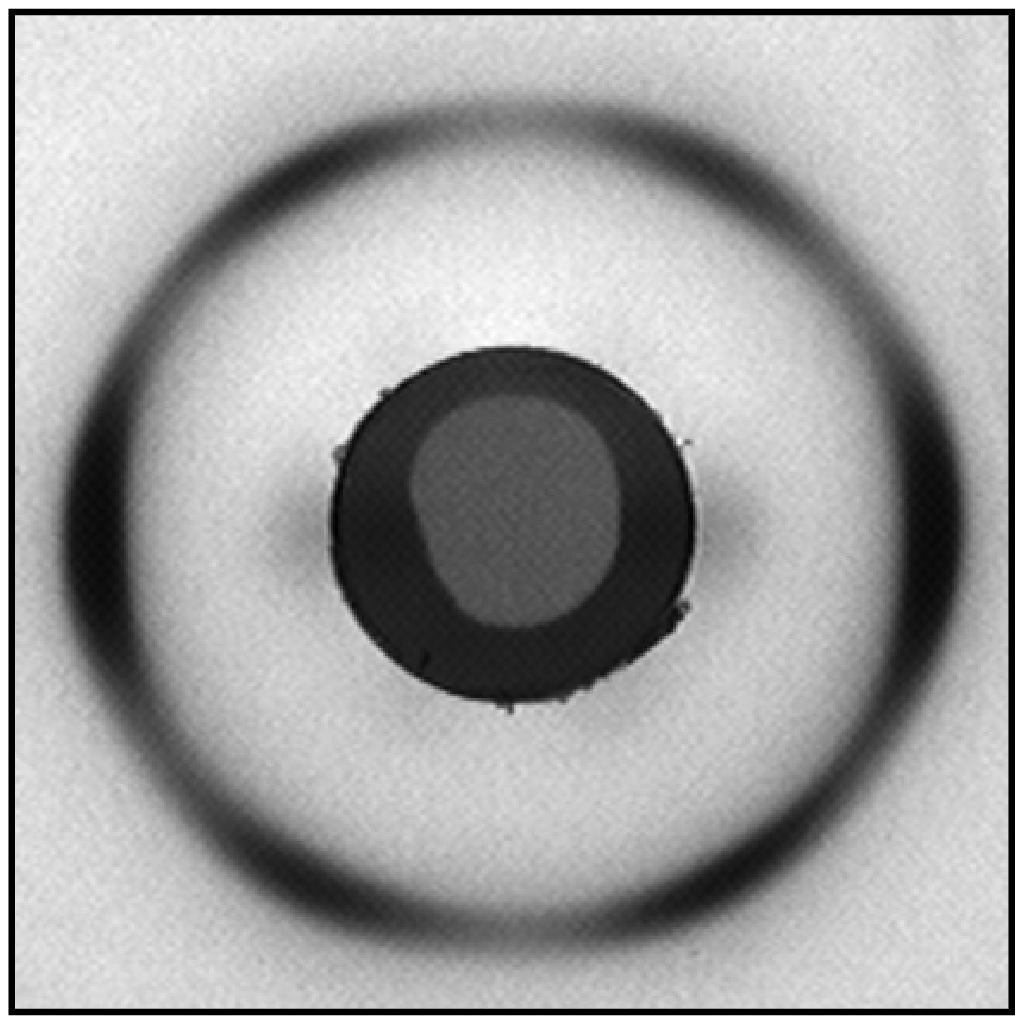}}}
\vskip 0.25truein
\centerline{\vbox{\baselineskip=0.20truein\hsize=4truein
\noindent FIGURE 1. X-ray structure function in the plane perpendicular
to the nematic direction\NIH.  This diffraction pattern contains
a non-zero $\cos 6\theta$ component and no measurable $\cos 6n\theta$ for $n\ge
2$.
The small amount of $\cos 2\theta$ can be attributed to the misalignment
of the X-ray beam.  (Figure provided courtesy of R.~Podgornik).
}}
\bigskip
If the chiral interactions are small, however,
the hexatic order may not twist.  Fortunately, there are reasons to
believe that this may be the case:
in the last section of this talk I will argue
that molecular chirality manifests itself as mesoscopic chirality ({\sl i.e.},
cholesteric pitch) via a subtle short-range correlation of molecular
orientation.
Na\"\i vely, one might expect that the cholesteric pitch should be on
the order of the molecular scale divided by a fraction of a radian, $P\approx
20\angstrom/(0.1\pi)$.  Typically pitches are {\sl much} longer, on the order
of microns.  The necessity of intramolecular correlations may shed some
light on the ``naturalness'' of observed cholesteric structures.
\oneskip\oneskip
\noindent\underbar{CHIRAL BOND ORDER AND EXPULSION OF CHOLESTERIC
TWIST}\oneskip

Based on symmetry let us construct a Landau free energy to describe the
phases of chiral molecules with nematic and hexatic phases\KN\nudge\LCBO.
In the non-chiral nematic phase, orientational fluctuations are
controlled via the Frank free energy density:
\eqn\efr{\free_n = {K_1\over 2}\left(\div\nb\right)^2 + {K_2\over 2}\left(
\nb\dot\curl\nb\right)^2 +{K_3\over
2}\left[\nb\times\left(\curl\nb\right)\right]^2.}
If there is hexatic order, there will also be a spin stiffness for
the hexatic bond-angle $\theta_6$.  When determining the free energy for
$\theta_6$, one must take into account the transformation properties
of $\theta_6$ under the nematic symmetry $\nb\rightarrow -\nb$.
Since angular changes must be measured with respect to some vector
(via a ``right-hand'' rule), under nematic inversion
$\theta_6\rightarrow -\theta_6$.
This does not change the sense of any twisting present in $\theta_6$
since the angular change is measured with respect to $\nb$.
Taking into account the local nematic anisotropy,
the elasticity for $\theta_6$ is:
\eqn\efri{\eqalign{\free_6 &= {K_A^{||}-K_A^{\perp}
\over 2}\left[\nb\dot\nabla\theta_6\right]^2
+{K_A^{\perp}\over 2}\left[\nabla\theta_6\right]^2\cr
&\qquad\qquad + K_6\curl\nb\cdot
\nabla\theta_6+
K_6'\left(\nb\dot\curl\nb\right)\left(\nb\dot\nabla\theta_6\right).\cr}}
These terms are invariant under the simultaneous change $\left(\nb,\theta_6
\right)\rightarrow -\left(\nb,\theta_6\right)$.  The combined
free energies have been considered before and constitute the elasticity
theory for the ``N+6'' phase\TON\nudge\GIA.

In a system composed of chiral molecules (such as DNA), one expects
additional chiral terms invariant under nematic inversion but
which change sign under spatial inversion.
Since nematic symmetry already forces all terms to have even powers of
$\nb$ and $\theta_6$, a character need not be assigned
to these variables ({\sl i.e.} vector or pseudovector,
pseudoscalar or scalar, respectively).  Thus I include the following
chiral
terms, quadratic in the fields:
\eqn\echi{\free^*=K_2q_0\nb\dot\curl\nb + K_A^{||}\tilde
q_0\nb\dot\nabla\theta_6.}
The first term is the usual chiral term which favors a cholesteric
texture $\nb=\left[\cos q_0z,\sin q_0z,0\right]$ while the second
term favors the twisting of the hexatic order\LCBO.  The second
term has been considered before in cholesteric melts\TER$\!$ and
chiral polymer crystals\KN .  Similar theories could be constructed
for molecules or molecular packings
with other sorts of in-plane symmetries, for instance
tetrahedral molecules\FEL.
\bigskip
\centerline{\vbox{\epsfxsize=4truein\hsize=4truein\epsfbox{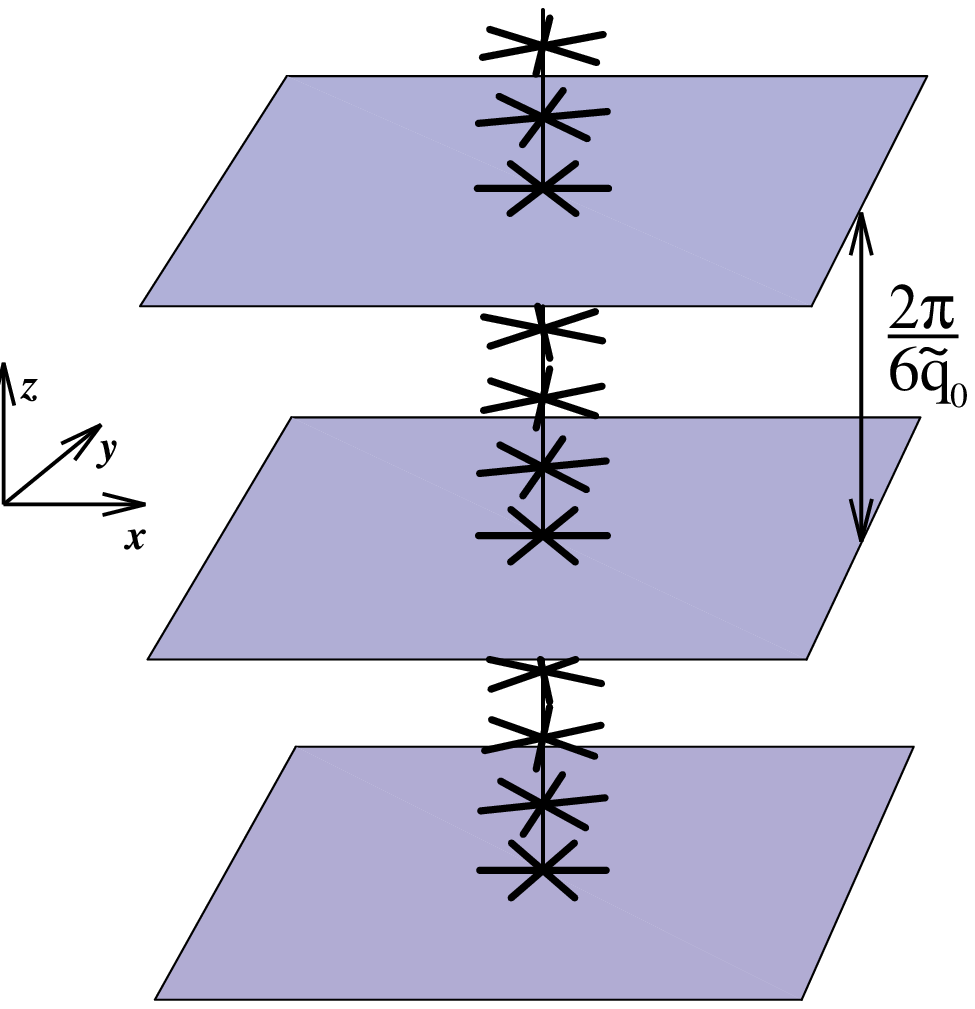}}}
\vskip 0.25truein
\centerline{\vbox{
\hsize=4truein\baselineskip=0.20truein
\noindent FIGURE 2. A chiral hexatic.  In each plane the bond-angle
order parameter is $\theta_6=\theta_6^0~{\rm mod}~2\pi/6$.  The
bond-angle rotates smoothly between the planes.  Each plane of constant
bond-angle is analogous to a smectic layer.  This phase has no density
modulations
}}
\bigskip
In the following I simplify to the case $K_A^{\perp}=K_A^{||}=K_A$,
$K_6=K_6' =0$ and employ the {\sl ansatz} first considered by Meyer\MEYER\ for
a nematic in a magnetic field:
$\nb=\left[\cos\phi\cos qz,\cos\phi\sin qz,\sin\phi\right]$, with
$\phi$ constant.  Minimizing
the total free energy
$F=\int d^3\!x\left[\free_n+\free_6
+\free^*\right]$
with respect to $\theta_6$ yields an effective free energy
for $\nb$ (note
that the mean-field equations only allow constant $\nabla\theta_6$):
\eqn\eeff{\free_{\rm eff} = \free[\nb;\tilde q_0]
- {K_A\over 2}\tilde q_0^2 \left(\hat z
\dot\nb\right)^2.}
This is precisely the free energy considered by Meyer for a nematic
with negative magnetic susceptibility in a magnetic field.  It is
straightforward to show that for $K_2<K_3$ the cholesteric phase
($\phi=0$) is the lowest energy state (within this class of solutions\ref\comm{
This solution also satisfies the Euler-Lagrange equations of mean-field
theory and so it is an extremum.}\ when $K_2q_0^2>K_A\tilde q_0^2$
and that the nematic phase ($\phi=\pi/2$) is lowest in energy
otherwise.  When $K_2>K_3$ an intermediate conical phase with
$0<\sin\phi<1$ is allowed\MEYER\nudge\LCBO$\!$, though typically this
inequality of the Frank constants is not satisfied.

We have just seen that the presence of hexatic order can act as
a field.  In fact, the chiral coupling to the hexatic bond-order field
makes this system identical to that of a smectic liquid crystal
composed of chiral mesogens with a layer spacing of $2\pi/(6\tilde q_0)$.
The hexatic ``wave'' plays the role of the smectic layers.  Each ``layer''
is a place where the bond-angle is $\theta_6=0~{\rm mod}~2\pi/6$.  This
is illustrated in Figure 2.  As in smectics, there will be
a myriad of phases, though, in analogy with a type-II superconductor\DG,
the simplest phases will be the cholesteric phase which has no chiral bond
order (normal
metal), the chiral hexatic phase with no cholesteric twist (Meissner phase),
and
finally the twist-grain-boundary (TGB)
phase\TGB\ (Abrikosov flux lattice phase).

This analogy can be made precise\LCBO\ by introducing the hexatic
order parameter $\psi_6$.  The Landau theory which describes the
liquid-to-hexatic transition is:
\eqn\ellh{\free_6 = \vert\nabla\psi_6\vert^2 + r\vert\psi_6\vert^2 +
u\vert\psi_6\vert^4,}
where $r\propto (T-T_c)$.  In the ordered phase ($T<T_C$)
$\psi_6=\vert\psi_6\vert
e^{i\theta_6}$ and
$\vert\psi_6\vert^2=K_A/72$.
Adding the second term of \echi\ as well as the Frank free energy for a nematic
with pitch $q_0$ ($\free[\nb;q_0]$) yields:
\eqn\elto{\free_6^* =\vert\nabla\psi_6\vert^2 + r\vert\psi_6\vert^2 +
u\vert\psi_6\vert^4
- 6i\tilde q_0\nb\dot\left(\psi^*\bold{\nabla}\psi
-\psi\bold{\nabla}\psi^*\right) + \free[\nb;q_0].}
Completing the square, this leads to:
\eqn\eltoo{\free_6^* =\vert\left(\bold{\nabla}+6i\tilde
q_0\nb\right)\psi\vert^2
+\left(r-36\tilde q_0^2\right)\vert\psi\vert^2 + u\vert\psi\vert^4 +
\free[\nb;q_0],}
which is precisely the free energy of a smectic-A liquid crystal with layer
spacing\DG\ $a=2\pi/(6\tilde q_0)$.

Drawing upon the smectic-A analogy, one would expect that cholesteric twist
will
be excluded in favor of {\sl chiral} hexatic order.  In particular, this means
that from plane to plane the sixfold order should rotate if there is
nematic alignment.  This is not what is seen experimentally -- there are
six distinct spots in the scattering perpendicular to the nematic director\NIH.
While the molecular symmetry suggests that the chiral terms in \echi\ must
be present, they may nonetheless be small.  In the next section I
will analyze the origin of chiral interactions among molecules and argue
that they could be anomalously small.  It should be pointed out that if
all the chiral terms were set to zero there is no {\sl a priori} reason why the
phase sequence nematic--N+6--hexagonal-columnar-phase could not
exist\ref\comtwo{A hexatic phase would not generically exist due
to the cubic invariants that could be constructed in Landau theory\KN.
Nonetheless, it is possible, in principle, for these cubic terms
to be tuned to zero and for the hexatic phase to exist.}.
\oneskip\oneskip
\noindent\underbar{THE MICROSCOPIC BASIS OF CHIRAL INTERACTIONS}\oneskip

A molecule is chiral if its symmetry group does not contain the
element $\bf S_n$:
a rotation around a $C_n$-axis by $2\pi/n$ followed by a mirror
through the perpendicular plane\ref\KELVIN{W. Thomson,
\underbar{The Robert Boyle Lecture}, (Oxford University
Junior Scientific Club, May 16, 1893) reprinted in the \underbar
{Baltimore Lectures on Molecular Dynam-} \underbar{ics and the Wave Theory
of Light},
(C.J. Clay \& Sons, London, 1904).}.  Moreover, all chiral
molecules must have $n$-axial order around some axis since the groups
${\cal C}_\infty$ and ${\cal D}_\infty$ are not allowed
for any real
objects\ref\LL{L.D. Landau and E.M. Lifshitz,
\underbar{Quantum Mechanics}, Third Edition (Pergamon Press, Oxford,
1977), Chap. XII.}: the
only infinite subgroups of the rotation group in three dimensions are
${\cal C}_{\infty v}$ and ${\cal D}_{\infty h}$, both of which contain ${\bf
S_1}$.
As we shall see, the lack of uniaxial symmetry plays a crucial role in
the chiral interactions.  Let us first concentrate on calculating the
cholesteric
pitch $q_0$, or, more accurately, $K_2q_0$ the coefficient
of $(\nb\cdot\curl\nb)$, and
return later to estimating $\tilde q_0$.
The following will only be a sketch of
the main results presented elsewhere\ref\HKL{A.B.~Harris, R.D.~Kamien
and T.C.~Lubensky, in preparation (1996).}.
To this end, let us consider two planes of
molecules separated along the $x$-axis by $R$, the first
aligned along the $\hat z$ axis, the second along $(\hat z\cos\theta+\hat
y\sin\theta)$.
Expanding for $\theta\ll 1$ yields:
\eqn\eto{F = \int d^3\!x\,K_2q_0\partial_x\delta n_y + \bo{\theta^2} \approx
K_2q_0L_yL_z\theta,}
where $L_y$ and $L_z$ are the sample dimensions.  There is a famous
expression for the coefficient of $\theta$ in this expansion--the torque.
Thus $L_yL_zK_2q_0$ is the torque that one plane of molecules exerts on the
next plane of molecules when they are aligned along a common axis.
In the simplest case, let us consider weakly interacting
molecules and thus build the torque up out of nearest neighbor, pairwise
interactions across the planes.  If there are $N$ molecules per plane then
the torque between two of the molecules is $AK_2q_0$, where $A$ is the free
area of each molecule.

We see that the problem of calculating $K_2q_0$ reduces to the problem of
calculating the torque between two molecules separated by a distance $R$
when they are aligned with each other and perpendicular to their separation.
For simplicity, let us assume that the atoms on the molecules all interact via
the same
central potential.  This analysis necessarily excludes quantum interactions
which can be non-local and which do produce chiral terms in the
Landau theory of liquid crystals\ref\KATS{
E.I. Kats, \underbar{Zh. Eksp. Teor. Fiz.}, \underbar{74}, 2320 (1978)
[\underbar{Sov. Phys.
JETP}, \underbar{47},
120 (1978)].}, but does include, as a limiting case, hard-core steric
interactions which can be built up out of short range central forces.
The relevant component of the torque is along $\Rbh\equiv\Rb/\vert\Rb\vert$,
where
$\Rb$ is the separation between the centers of mass
of each molecule and $\Rb\cdot\nb=0$.  The atoms on each molecule are at a
position $\rb_\alpha$ relative to the center of mass and
$\alpha$ runs over the atoms on the molecule.  Assuming a central
potential $V(\Rb)$, the relevant component of the torque is:
\eqn\ecalctorque{
\Rbh\cdot\bold{\tau} = -
{1\over\vert\Rb\vert}\sum_{\alpha\beta}
R^i\epsilon_{ijk}r^j_\beta\partial^kV(\Rb+\rb_\beta
-\rb_\alpha),}
where $\alpha$ runs over the atoms on the first molecule
and $\beta$ runs over the atoms on the second molecule.
An expansion in powers of $(\vert\rb\vert/\vert\Rb\vert)$
is certainly reasonable in
\bigskip
\centerline{\vbox{\epsfxsize=4truein\hsize=4truein\epsfbox{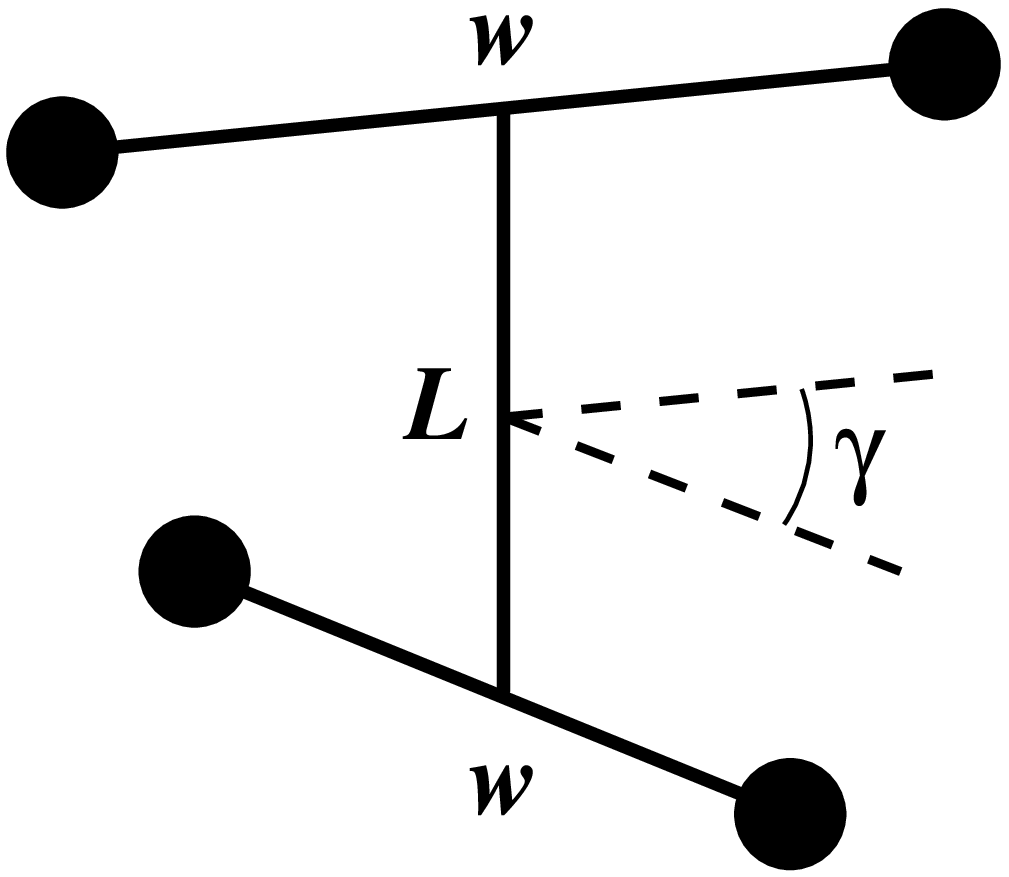}}}
\vskip 0.25truein
\centerline{\vbox{\hsize=4truein\noindent FIGURE 3. Prototypical chiral
molecule.  The spheres indicate
the positions of the atoms.  When the angle $\gamma$ is $n\pi/2$ this
molecule is not chiral.  The chiral order parameter for this molecule
is $\psi=-(1/8)w^4L\sin(2\gamma)$, where $L$ and
$w$ are the dimensions shown and $L>w$ (not drawn to scale).
}}
\bigskip
\noindent the dilute limit and the
general symmetry of the interaction should not be significantly
altered at higher densities.  Note that terms in the expansion
which are {\sl even} in $\rb$ must vanish:  under $\rb\rightarrow -\rb$ both
molecules will change their handedness and thus the chiral pitch will change
sign.  Note that $\partial^kV(\Rb)\propto R^k$
and thus from the antisymmetry of $\epsilon_{ijk}$ the first order term
will vanish.  The third order term is:
\eqn\ethior{\left(\Rbh\cdot\bold{\tau}\right)_3
= -{1\over 2!}{1\over\vert\Rb\vert}
\sum_{\alpha\beta} R^i\epsilon_{ijk}r_\beta^j\left(r_\beta^m-r_\alpha^m\right)
\left(r_\beta^l-r_\alpha^l\right)\partial^k\partial^m\partial^lV(\Rb).}
The potential term $\partial^k\partial^m\partial^lV(\Rb)$ will, by
symmetry, have only two types of terms: those proportional to 1) $R^kR^mR^l$
and
2) $\left(R^k\delta^{ml} + R^m\delta^{kl} +R^l\delta^{km}\right)$.  Again,
due to the antisymmetry of $\epsilon_{ijk}$,
terms of the first type will vanish and
the only part of \ethior\ which survives is:
\eqn\ethiorr{\left(\Rbh\cdot\bold{\tau}\right)_3
\propto \sum_{\alpha\beta} R^i\epsilon_{ijk}r_\beta^jr_\alpha^k
\Rb\cdot\left(\rb_\beta-\rb_\alpha\right).}
Note that every term in \ethiorr\ will contain the single
sum $\sum_\mu r^i_\mu$ which is identically $0$
since the $\rb_\mu$ are measured from the center of mass.  Thus, the
third order term will vanish.

Fortunately, perseverance will pay off and the fifth order term
will not identically vanish.  At fifth order there are a variety of
terms.  Again, any term which is a sum over a single factor
of $r^i_\mu$ will vanish.  This leaves only terms with three $r^i$ on
one molecule and two $r^j$ on the other.  Taking the average
yields\HKL:
\eqn\efifor{\left(\Rbh\cdot\bold{\tau}\right)_5
=U(R)Q^{il}\epsilon_{ijk}\left\{S^{jlm}_2B_1^{km} + S^{jlm}_1B_2^{km}\right\},}
where $U(R)=\{{1\over 8}R^3f^{(4)}(R^2/2)-{1\over 2}Rf^{(3)}(R^2/2)\}$ is a
function of $R=\vert\Rb\vert$,
$V(R)=f(R^2/2)$,
$Q^{il}\equiv\left[n^in^l-(1/3)\delta^{il}\right]$ is the nematic alignment
tensor (equal for both molecules), the two tensors
$S_p^{jlm}$ and $B_p^{km}$ are defined on molecule $p=1,2$ via
\eqn\etensdef{\eqalign{
S_p^{jlm} &\equiv \sum_{\mu\in p} \left\{r^j_\mu r^l_\mu r^k_\mu - {1\over
9}\vert\rb_\mu\vert^2\left[
\delta^{jl}r^k_\mu +\delta^{jk}r^l_\mu +\delta^{kl}r^j_\mu\right]\right\}
\cr
B_p^{km} &\equiv \sum_{\mu\in p} r^s_\mu r^t_\mu\left
\{\tilde\delta^{sk}\tilde\delta^{tm} - {1\over
2}\tilde\delta^{km}\tilde\delta^{st}\right\},
\cr}}
and $\tilde\delta^{ij} =\left[\delta^{ij} - n^in^j\right]$ is
the transverse projection operator.
The tensor $B^{km}$ measures the biaxial orientation of the molecules in
the plane perpendicular to the nematic direction, which is assumed to be
the principal
axis of the molecule with the largest moment.  The other tensor $S^{jlm}$
is some measure of the chirality of the molecule, but is not a measure by
itself:
indeed, some components of $S^{jlm}$ will be non-zero for achiral molecules and
zero for chiral molecules.
Consider, however,
the order parameter $\psi_p = S_p^{jlm}\epsilon_{ijk}Q^{il}B_p^{km}$
on a single molecule $p$: in the co\"ordinate system of the
principal axes, $\psi = (\lambda_y-\lambda_x)
\sum_\mu r^x_\mu r^y_\mu r^z_\mu$ where $\lambda_x$ and $\lambda_y$ are
the eigenvalues along the $\hat x$ and $\hat y$ directions of the
single-molecule
moment of inertia tensor and $\hat z$ is the
nematic axis.  This order parameter vanishes identically for achiral
molecules, since under inversion the principal axes do not change but the
sum will change sign.  Solving for
$S^{jlm}_p = \psi_p\epsilon_{ijk}Q^{il}B^{km}_p/(2B_p^2)
+ (\hbox{symmetric permutations})$,
\efifor\ reduces to:
\eqn\efiforr{
\left(\Rbh\cdot\bold{\tau}\right)_5 = \psi\,U(R){B_1^{km}B_2^{km}\over B^2}.}
where $B^2=B_1^2=B_2^2$ is the same on the two identical molecules.
Up to this point, the average over the $n$-axial direction
has not been performed.
In any mean field theory $\langle\,B^{km}\,\rangle$ will either be zero
(uniaxial)
or non-zero.  In the latter case the entire phase will be {\sl biaxial}, with
long-range biaxial order.  If the same mesogens in racemic
mixtures have only uniaxial nematic phases then $\langle\,B^{km}\,\rangle=0$
({\sl i.e.}, no long-range biaxial order).
Thus if the molecules rotate
{\sl independently} around their nematic axes then
the average $\langle\,B^{km}_1B^{km}_2\,\rangle=\langle\,B^{km}_1\,\rangle
\langle\,B^{km}_2\,\rangle=0$.
Thus in a uniaxial phase,
chiral interactions will come about only via {\sl short-range} correlations of
biaxial order.  In other words, upon averaging, \efiforr\ becomes
\eqn\efiforrr{
\left\langle\,\Rbh\cdot\bold{\tau}\,\right\rangle_5 =
\psi\,U(R){\langle\,B_1^{km}B_2^{km}\,\rangle\over B^2}
\equiv \psi\,U(R)g_B(R),}
where $g_B(R)\sim e^{-R/\xi_B}$ is the normalized biaxial correlation function,
and $\xi_B$ is the biaxial correlation length,
which one might na\"\i vely expect to be on the scale of the intramolecular
spacing.

The previous discussion shows that there are two factors which lead to
a large chiral parameter $q_0$:  the strength of the chiral order parameter
$\psi$
on the molecule and the range of the biaxial correlations.  If either is small,
it
is reasonable to expect a very long pitch.  Note that almost cylindrical,
corkscrew-like
molecules are {\sl not} very biaxial and thus $\psi$ would be small.
In addition, in the DNA system studied at NIH, the molecules are on
the order of $40\angstrom$ apart (center-to-center) and are each roughly
$20\angstrom$
wide.  Moreover, the Debye screening length at the typical
salt concentrations investigated is
on the order of $5\angstrom$ and thus one might expect
that the
biaxial interactions would be weak and hence $\xi_B$ to be small.
Together, the smallness of both
factors suggests that the chirality might be very small.  In fact, it is
even possible that in the hexatic phase biaxial correlations are
{\sl smaller} than in the cholesteric phase, despite the higher molecular
density.  Since the cholesteric phase picks out a preferred, biaxial direction
(the pitch axis) the biaxial correlations could be longer-ranged (even
possibly long-range).  In a hexatic phase, however, the biaxial order
parameter might be suppressed due to the $6$-fold ``crystal'' field of
the hexatic.  Note that if the biaxial correlations are removed there
could still be ``$6$-axial'' correlations.
In that case, however, the analogous
order parameter to $\psi$ would be more complicated and would include
higher geometric moments.  Moreover, the expansion of $\Rbh\cdot\bold{\tau}$
in powers of $(\vert\rb\vert/\vert\Rb\vert)$ would vanish, upon averaging, even
at fifth
order, and one would have to go to, at least, {\sl ninth} order
which would lead
to a very small interaction at dilute separations.

Finally, let us estimate the amount of twisting of the hexatic order
by considering a simplified two-polymer model.
Consider one polymer
helically winding its way around another straight polymer at a mean separation
$a$:
the rate at which the separation vector rotates
along the straight polymer axis is a simple estimate of $\tilde q_0$ and
will be precisely equal to the pitch of the helical polymer trajectory.
There are two major contributions to the energy of the helical polymer:
the first is the usual bending energy of a stiff
polymer, while the second is the ``nematic'' free energy
for the two molecules which includes the cholesteric term proportional
to $q_0$.
Take
the first polymer to follow the curve $\rb(z)=\left[a\cos\tilde q_0z,
a\sin\tilde q_0z,z\right]$ and the second to lie along the $\hat z$-axis
at $x=y=0$.  In this case $\nb\cdot\curl\nb
\approx (a\tilde q_0)/a=\tilde q_0$, and so
the free energy of the second polymer is:
\eqn\epoly{F = \int dz\,\left\{{\kappa\over 2}\left\{1+(a\tilde
q_0)^2\right\}^{-3/2}
a^2(\tilde q_0)^4 + {K_2\over 2}a^2\left[\tilde q_0 +q_0\right]^2\right\}.}
Minimizing for small $a\tilde q_0$ and taking the bending stiffness
$\kappa=\kbT L_P$, where $L_P$ is the persistence length, yields:
\eqn\esolve{ 2L_P\left(\tilde q_0\right)^3 + {K_2\over\kbT}\tilde q_0
= - {K_2\over\kbT}q_0.}
Taking the typical DNA values for
$K_2 = 10^{-6}$ {\sl dyne}, $q_0 = 2\pi/\mu m$
and $L_P=600\angstrom$,
I expect that at room temperature $\tilde q_0\approx -q_0$.  It
should be noted, however, that as $L_P\rightarrow\infty$, $\tilde q_0
\rightarrow 0$, in accord with one's intuition -- as the polymers
become infinitely stiff there can be no hexatic twisting. More
generally, the elastic constant $K_2$ has been estimated for
semi-flexible polymers with mean spacing $a$ (roughly $40\angstrom$
in the DNA samples) to
be\ref\SB{J.V.~Selinger and R.F.~Bruinsma, \underbar{Phys. Rev. A},
\underbar{43}, 2910 (1991).}\ $K_2\approx \kbT L_P^{1/3}/a^{4/3}$.
Substituting into \esolve\
I find $\tilde q_0\sim -q_0$ for pitches $P=2\pi/\tilde q_0
> 2\pi\left(L_Pa^2\right)^{1/3} \approx 600\angstrom$, which is almost
always the case.
Thus, because the typical pitches are much longer than any intrinsic
molecular
length, the polymer stiffness plays no significant role in
reducing $\tilde q_0$.
\vfill\eject
\noindent\underbar{ACKNOWLEDGEMENTS}\oneskip
Some of this work was done in collaboration with A.B.~Harris and T.C.~Lubensky.
I thank them for allowing me to present some of our results here.
It is a pleasure to acknowledge stimulating discussions with them, as well
as with S.~Fraden,
R.~Meyer, D.~Nelson, P.~Nelson,
V.A.~Parsegian, R.~Pelcovits, R.~Podgornik, H.~Strey and J.~Toner.
This work was supported in part by the National Science Foundation,
through Grant No.~DMR94-23114.

\footatend\bigskip\immediate\closeout\rfile\writestoppt
\baselineskip=.20truein\leftline{\underbar{REFERENCES}}\bigskip{\frenchspacing%
\parindent=20pt\escapechar=` \input refs.tmp\vfill\eject}\nonfrenchspacing

\bye